\documentclass[aps,epsfig,twocolumn]{revtex4}
\usepackage{amsmath}
\usepackage{epsfig}
\begin{document}
\title{Two-component Fermi gas with a resonant interaction}
\author{G.\ M.\ Bruun}  
\affiliation{Niels Bohr Institute, Blegdamsvej 17, 2100 Copenhagen, Denmark}
\begin{abstract} 
We consider a two-component Fermi gas interacting via a Feshbach molecular state. 
It is shown that an  important energy scale is $E_g=g^4m^3/(64\pi^2)$ where 
$g$ is the Feshbach coupling constant and $m$ the mass of the particles. 
 Only when $E_g\gg \epsilon_{\rm F}$ where $\epsilon_{\rm F}$ is the Fermi energy 
can the gas be expected to  enter a universal state in the unitarity limit 
on the atomic side of the resonance 
where there are no molecules present. The universal state is distinct from the molecular 
gas state on the other side of the resonance. We furthermore calculate the energy of the gas 
for this universal state and our results are related to 
current experiments on $^{6}$Li and $^{40}$K.
\end{abstract}
\maketitle

One of the most interesting recent developments within the field 
of ultracold atom gases is the use of Feshbach resonances to 
manipulate the effective atom-atom interaction. By controlling an 
external magnetic field, the Feshbach molecular state can be tuned 
to an energy close to the scattering atoms. In this case, the vacuum
atom-atom scattering is unitarity limited  with a diverging scattering 
length $a$ and the system is strongly interacting. Thus, it 
is possible with the atomic gases to 
address the important question  of the nature of a many-body system 
with an interaction characterized by $|a|\rightarrow\infty$.
Indeed, remarkable experimental progress concerning this question has been 
made in recent months with several groups  reporting results 
for two-component atomic Fermi ($^6$Li and $^{40}$K) gases interacting via a 
closed channel Feshbach molecular state close to 
resonance~\cite{Duke,ENS,JILAMIT,Innsbruck}.

In this paper, we examine the 
problem of a two component Fermi gas interacting via a Feshbach molecular state 
close to zero energy. We show that only for sufficiently broad resonances 
with $E_g\gg \epsilon_{\rm F}$ can the gas be expected to exhibit universal behavior 
close to resonance on the $a<0$ when there are no molecules.
 The possibility of such universal behavior when 
 $|a|\rightarrow\infty$ has been examined by a number of 
authors~\cite{Heiselberg,Ho,Baker,Carlson}. 
The basic idea is that $|a|$ is not a relevant length scale in this limit and that the
thermodynamic quantities of the gas
therefore should only depend on the density $n$ and the temperature $T$. The gas is 
in this sense universal since its properties are independent of the specific details 
of the interaction. The energy per particle 
for a Fermi gas with  $T\ll T_{\rm F}$ where $T_{\rm F}$ is the Fermi temperature
is predicted to scale as $E/N=3\epsilon_{\rm F}(1+\beta)/5$ in the universal limit 
 and various values for the constant $\beta$ has been predicted~\cite{Heiselberg,Baker,Carlson}.
For a Bose gas with $|a|\rightarrow\infty$, it was argued that $E/N\propto n^{-2/3}$~\cite{Cowell}. 
Similar results where derived for bosonic three particle states~\cite{Jonsell}. 
We will in the present paper calculate $\beta$ for a Fermi gas in the universal limit for 
$E_g\gg \epsilon_{\rm F}$ taking into account two-body scattering in a medium. 
Our results are finally related to the present experiments on the atomic gases.

The explicit model considered here consists 
of atoms with mass $m$ in two hyperfine states denoted $\uparrow$
and $\downarrow$ being coupled to a closed channel with a Feshbach molecule. 
The Hamiltonian describing this interacting mixture  of atoms and molecules 
(described as point bosons) is
\begin{gather}
\hat{H}=\hat{H}_0+
\sum_{{\mathbf{K}},{\mathbf{q}}}\frac{g_{\rm bare}(q)}{\sqrt{\mathcal{V}}}
[\hat{b}^\dagger_{\mathbf{K}}\hat{a}_{{\mathbf{K}}/2+{\mathbf{q}}\uparrow}
\hat{a}_{{\mathbf{K}}/2-{\mathbf{q}}\downarrow}+ {\rm h.c}]
\nonumber\\
+\sum_{{\mathbf{k}},{\mathbf{q}},{\mathbf{q}}'}\frac{V({\mathbf{q}},{\mathbf{q}}')}{\mathcal{V}}
\hat{a}^\dagger_{{\mathbf{k}}+{\mathbf{q}}\uparrow}
\hat{a}_{{\mathbf{k}}-{\mathbf{q}}\downarrow}^\dagger\hat{a}_{{\mathbf{k}}-{\mathbf{q}}'\downarrow}
\hat{a}_{{\mathbf{k}}+{\mathbf{q}}'\uparrow}
\nonumber
\end{gather}
with 
$
\hat{H}_0=\sum_{{\mathbf{k}},\sigma}\epsilon_k\hat{a}^\dagger_{{\mathbf{k}}\sigma}\hat{a}_{{\mathbf{k}}\sigma}+
\sum_{\mathbf{K}}E_K^{\rm bare}\hat{b}^\dagger_{\mathbf{K}}\hat{b}_{\mathbf{K}}
$. 
 Here $E_K^{\rm bare}=2\nu_{\rm bare}+K^2/4m $, where $2\nu_{\rm bare}$ is the energy of a bare
molecule with momentum zero 
measured with respect to the energy of a pair of atoms at rest in the open
channel, $\epsilon_k=k^2/2m$ is the kinetic energy of an atom, 
$g_{\rm bare}(q)$ is the bare molecule-atom coupling matrix element, $V$ is the non-resonant 
interaction between atoms, and ${\mathcal{V}}$ is the volume of the system.

The main object to calculate here is the thermodynamic potential for the gas 
$\Omega=-k_BT\ln Z$ with $Z=\rm{Tr}\{exp[-\beta(\hat{H}-\mu\hat{N}_{\rm F}-2\mu\hat{N}_B)]\}$.
Here,  $\hat{N}_{\rm F}=\sum_{{\mathbf{k}},\sigma}\hat{a}^\dagger_{{\mathbf{k}}\sigma}\hat{a}_{{\mathbf{k}}\sigma}$ and 
$\hat{N}_B=\sum_{\mathbf{K}}\hat{b}^\dagger_{\mathbf{K}}\hat{b}_{\mathbf{K}}$
are the number of bare atoms and molecules respectively and $\mu$ is the chemical potential. 
We assume an equal density of the two atomic hyperfine states and a 
total density $n=k_{\rm F}^3/3\pi^2$. Also, the molecules and atoms are assumed to be in equilibrium.
Using the ladder approximation (or the Gaussian approximation for the partition function),
 the thermodynamic potential can after some algebra be expressed as 
\begin{gather}
\frac{\Omega}{\mathcal{V}}=\frac{\Omega_0}{\mathcal{V}}+
\int\frac{d^3K}{(2\pi)^3}\int\frac{d\omega}{\pi}\frac{1}{e^{\beta\omega}-1}\times\nonumber\\\rm{Im}\,
\rm{Tr}\ln[1-(V+g_{\rm bare}D_0(K,\omega)g_{\rm bare})G_0^{(2)}({\mathbf{K}},\omega)].
\label{Thermo}
\end{gather}
Here $\Omega_0$ is the thermodynamic potential of an ideal  gas of non-interacting atoms and molecules
 described by 
$\hat{H}_0$. We omit for brevity here and in the rest of the paper a small positive imaginary part to 
the frequencies 
in the (retarded) Greens functions which should really be evaluated at $\omega+i\delta$. 
The free molecule Greens function  is $D_0(K,\omega)^{-1}=\omega-E_K^{\rm bare}+2\mu$ and the two particle 
Greens function 
$G_0^{(2)}({\mathbf{K}},{\mathbf{q}},\omega)=[f(\xi_{{\mathbf{K}}/2+{\mathbf{q}}})+
f(\xi_{{\mathbf{K}}/2-{\mathbf{q}}})-1]/[\xi_{{\mathbf{K}}/2+{\mathbf{q}}}+
\xi_{{\mathbf{K}}/2-{\mathbf{q}}}-\omega]
$
describes the free propagation of a pair of atoms with COM momentum ${\mathbf{K}}$ and relative 
momentum ${\mathbf{q}}$ in the presence of the medium.  
Here $\xi_k=\epsilon_k-\mu$ and $f(x)=[\exp(x/k_BT)+1]^{-1}$ is the Fermi function. 
 The trace in  Eq.\ (\ref{Thermo}) refers to integration over the momentum  variables appearing in the matrices
 $V({\mathbf{q}},{\mathbf{q}}')$ and $g_{\rm bare}(q)G_0^{(2)}g_{\rm bare}(q')$. 
Various forms of Eq.\ (\ref{Thermo}) have been discussed in the literature in connection 
with the BCS-BEC crossover problem~\cite{Nozieres}.
For the present purposes, it is convenient to rewrite Eq.\ (\ref{Thermo}) 
as $\Omega=\Omega_0+\Delta\Omega$ with 
\begin{gather}
\frac{\Delta\Omega}{\mathcal{V}}=
\int\frac{d^3K}{(2\pi)^3}\int\frac{d\omega}{\pi}
\{\rm{Im}\rm{Tr}\ln[G_0^{(2)}({\mathbf{K}},\omega)/G^{(2)}_{\rm bg}({\mathbf{K}},\omega)]
\nonumber\\
+\rm{Im}\rm{Tr}\ln[D_0(K,\omega)/D(K),\omega)]\}(e^{\omega/k_BT}-1)^{-1}
\label{Thermo2}
\end{gather}
where ${G^{(2)}_{\rm bg}}^{-1}={G^{(2)}_{0}}^{-1}-V$ describes  
the propagation of a pair of atoms interacting only via the 
background interaction in the ladder approximation. 
$D^{-1}=D_{0}^{-1}-\Pi$ 
is the molecule Greens function with the self energy 
$\Pi(K,\omega)=\rm{Tr}[g_{\rm bare}G^{(2)}_{\rm bg}(K,\omega)g_{\rm bare}]$~\cite{bruunpethick}. 
Equation (\ref{Thermo2}) is convenient since it separates the corrections to $\Omega$
coming from the Feshbach molecule 
from the effects of the background (non-resonant) interaction. Since we focus in the effects of 
the resonant part of the interaction, we will 
in the rest of this paper only consider the molecule term in Eq.\ (\ref{Thermo2}). 
The $\ln[D_0(K,\omega)]$ term in Eq.\ (\ref{Thermo2}) is trivial as it simply cancels the bare
 molecule contribution to 
$\Omega_0$.

In order to be able to express our result in terms of observables only, 
we use the low energy effective theory developed in ref.~\cite{bruunpethick}. We therefore write 
$
\tilde{z}D^{-1}(K,\omega)=\omega-E_K+ig^2m^{3/2}(4\pi)^{-1}\sqrt{\tilde{\omega}}
-\tilde{\Pi}(K,\omega)
$ 
with  $E_K=K^2/4m+2\nu-2\mu$ and $\tilde{\omega}=\omega+2\mu-K^2/4m$.
 Here $2\nu$ is the energy of the molecule taking into account 
the high energy dressing by the atoms 
but excluding the threshold effects which are explicitly treated by the $\sqrt{\tilde{\omega}}$ 
term. The wave function renormalization of the molecule given by $\tilde{z}$ is  due to the coupling to 
high energy atoms. The constant $g$ 
yields the coupling  between this dressed molecule and the atoms 
and $\tilde{\Pi}$ gives the medium corrections 
to the molecule self energy $\Pi$. 
Explicit expressions for these quantities are given in ref.~\cite{bruunpethick} and will 
not be repeated here. The point is that $g$ and $2\nu$ can be extracted from experiments 
without any arbitrary assumptions of the behavior  of the bare quantities. 

It is easy to show that Eq.\ (\ref{Thermo2}) reproduces the two-body term in the 
virial expansion of $\Omega$ in 
the Boltzmann regime~\cite{Huang}. This expansion has been used to prove that 
a gas of atoms interacting via a closed channel resonant (zero energy) Feshbach molecule 
exhibits universal behavior in the non-degenerate limit~\cite{Ho}. 
We will presently  examine under which conditions such universal 
behavior persists in the low temperature degenerate regime.

We split $\Delta\Omega$ into a contributions coming from undamped and  damped two particle states.
This is done by writing  Eq.\ (\ref{Thermo2}) (neglecting the background parts) as
\begin{gather}
\frac{\Delta\Omega}{\mathcal{V}}=\frac{\Delta\Omega_{\rm M}}{\mathcal{V}}+
\int\frac{d^3Kd^3q}{(2\pi)^6}
\frac{f(\xi_{{\mathbf{K}}/2+{\mathbf{q}}})f(\xi_{{\mathbf{K}}/2-{\mathbf{q}}})}
{\rm{Im}\Pi(K,\xi_{{\mathbf{K}}/2+{\mathbf{q}}}+\xi_{{\mathbf{K}}/2-{\mathbf{q}}})/g^2}\times\nonumber\\
\arctan\left[\frac{\rm{Im}\Pi(K,\xi_{{\mathbf{K}}/2+{\mathbf{q}}}+\xi_{{\mathbf{K}}/2-{\mathbf{q}}})}
{q^2/m-2\nu-\rm{Re}\tilde{\Pi}(K,\xi_{{\mathbf{K}}/2+{\mathbf{q}}}+\xi_{{\mathbf{K}}/2-{\mathbf{q}}})}\right].
\label{Thermo3}
\end{gather}
where $\Delta\Omega_M=k_BT\sum_{\mathbf{k}}\log[1-\exp(\omega_K/k_BT)][1-\partial_w\tilde{\Pi}(k,\omega_K)]$
is the contribution from the undamped molecular states with energy $\omega_K$. 
The energies $\omega_K=\tilde{\omega}_K+K^2/4m-2\mu$ are determined by 
$\tilde{\omega}_K=2\nu+g^2m^{3/2}(4\pi)^{-1}\sqrt{-\tilde{\omega}_K}+\tilde{\Pi}(K,\omega_K)$
with $\omega_K\le K^2/4m-2\mu$ since $\rm{Im}\Pi(K,\omega)\neq 0$ for $\omega>K^2/4m-2\mu$. 
It is easy to show that $\tilde{\Pi}(K,\omega)\ge 0$
 for $\omega\le K^2/4m-2\mu$, i.e.\ the medium only increases the molecule energies. There will 
therefore be no undamped molecules below the Fermi sea for $\nu>0$.
To arrive at the second term in Eq.\ (\ref{Thermo3}) from Eq.\ (\ref{Thermo2}),
we have divided and multiplied by $\rm{Im}\Pi$
and performed the $\omega$ integration. Using Eq.\ (\ref{Thermo3}), we can analyze under which 
conditions the interacting gas will exhibit universal thermodynamics.

As explained above, the universal thermodynamics hypothesis states that 
the thermodynamic quantities should depend only on $T$ and $n$ in the unitarity 
limit when $|a|\rightarrow \infty$. 
For the specific system considered here, 
the scattering length is $4\pi a_{\rm res}/m=-g^2/2\nu$.
The  thermodynamic potential depends therefore in general on $\mu,\nu,T$,  and $g$ and 
we write $\Omega(T,\mu,\nu,g)$. If such a  gas is to exhibit 
universal thermodynamic behavior, we must 
have $\Omega(T,\mu,\nu,g)\simeq \Omega(T,\mu)$ close 
to resonance, i.e.\ the thermodynamics become independent 
of the specific details of the resonance. 

Let us first briefly consider the molecular side of the resonance with $\nu\le0$. 
Since $\mu\le\nu$, the system in this case consist of a pure molecular gas with essentially no atoms
present for $T\ll T_{\rm F}$ and we need only to consider $\Delta\Omega_M$ in Eq.\ (\ref{Thermo3}). 
We therefore also have $\tilde{\Pi}=0$ and the molecule energies are given by 
$\tilde{\omega}=-(\sqrt{E_g-2\nu}-\sqrt{E_g})^2$. We have introduced the energy 
$E_g=g^4m^3/(64\pi^2)$ which determines the range of energies for which threshold effects are important.  
So universality for $\nu\le0$ is straightforward to  understand within the present model:
 For $-E_g\ll 2\nu\le0$,
we simply obtain an ideal gas of molecules with energy $\tilde{\omega}=-\hbar^2/(ma_{\rm res}^2)$. 
However, this  result ignores the molecule-molecule interactions, which are not included in the
present formalism. Such effects could be important for $\nu\le 0$   
where there is a macroscopic number of molecules~\cite{Kokkelmans}.

The main purpose of the present paper is to examine the $a_{\rm res}<0$ ($\nu>0$) side of the resonance
where there are no undamped molecules and $\Delta\Omega_M$ 
in Eq.\ (\ref{Thermo3}) can be ignored. The only place where 
$g$ and $\nu$ enter in the expression for $\Omega$ is then in the argument of 
the $\arctan(\ldots)$ where $\nu$ 
appear explicitly and $g^2$ enters since $\Pi\propto g^2$. The molecule self energy 
 $\Pi$ scales as $\sim {\mathcal{O}}(g^2mk_{\rm F}/4\pi)$~\cite{bruunpethick}.  From this it follows that
in order to make the
argument of $\arctan(\ldots)$ independent of $g^2$ and $\nu$, we must have 
$g^2mk_{\rm F}/4\pi\gg \epsilon_{\rm F}$. Only in this case will the $\Pi$ terms in the 
 $\arctan(\ldots)$ dominate and thus cancel the $g^2$ and $\nu$ dependence 
even when $\nu\gg \epsilon_{\rm F}$. The reason for the requirement  $\nu\gg \epsilon_{\rm F}$ 
is that $\mu\le \nu$ from particle conservation. When  $\nu\lesssim \epsilon_{\rm F}$
we have $\mu\sim \nu$ due to the presence of damped molecules; i.e.\ $\mu$ and the thermodynamics 
of the gas will depend critically on $\nu$ when $0\le \nu \lesssim \epsilon_{\rm F}$. 
It should be
noted that for $\nu/\epsilon_{\rm F}\rightarrow 0$, we of course approach the pure molecular gas state for 
$\nu\le0$ described above; the state of the gas changes continuously across the resonance. 
What we have shown here however is that the gas approaches a \emph{different} 
universal state independent of $(g,\nu)$ for $\nu\gg \epsilon_{\rm F}$ 
 only if the resonance is sufficiently broad with $g^2\gg 4\pi k_{\rm F}m^{-2}$. This state is characterized 
by strongly correlated atom pairs with no real molecules present. 
The condition for universality on the $a_{\rm res}<0$ side can also be written 
\begin{equation}\label{Condition}
E_g\gg \epsilon_{\rm F}.
\end{equation}
In terms of the scattering length $4\pi a_{\rm res}/m=-g^2/2\nu$, Eq.\ (\ref{Condition}) can be written as
$k_{\rm F}|a_{\rm res}|\gg 1$ even when $ \nu\gg \epsilon_{\rm F}$. Thus, 
the condition for universality on the atomic side of the resonance simply states that the coupling 
must be so strong that one can reach the unitarity regime 
 $k_{\rm F}|a_{\rm res}|\gg 1$ before there is a significant 
population of molecules. The $q^2$ in the denominator of the $\arctan(\ldots)$ may be regarded 
as arising from an effective range term.  
We can then identify the effective range for the effective interaction mediated by the Feshbach molecule 
as $r_{\rm eff}=-8\pi m^{-2}g^{-2}$. Using this, we can finally write 
Eq.\ (\ref{Condition})  in the illuminating way $k_{\rm F}|r_{\rm eff}|\ll 1$, i.e.\ 
universality  can only be reached if the effective range of the interaction is small.

We now consider the $a_{\rm res}<0$ and $T=0$  limit of the universal 
behavior for  in more detail. 
For $T=0$, we can write 
$\Delta\Omega/{\mathcal{V}}=\mu n_{\rm id}(\mu)I$
where $I(\nu/\mu, g^2m^{3/2}/\sqrt{\mu})$ is the $T=0$ integral in Eq.\ (\ref{Thermo3}) 
expressed in dimensionless variables ($\tilde{k}=k/\sqrt{2m\mu}$ etc.) and 
$n_{\rm id}(\mu)=(2m\mu)^{3/2}(3\pi^2)^{-1}$.
 The integral $I$ can be evaluated numerically. In the universal 
limit $E_g\gg \epsilon_{\rm F}$, we get
$I_{\rm uni}\simeq -0.3$ for $T=0$. This 
immediately gives $\Omega(T=0,\mu)=-2\mu n_{\rm id}(\mu)(1-5I_{\rm uni}/2)/5$ and
 $\mu=(2m)^{-1}(3\pi^2n)^{2/3}(1-5I_{\rm uni}/2)^{-2/3}$.  
We finally obtain $E/N=3\epsilon_{\rm F}(1+\beta)/5$ with 
\begin{equation}
\beta=(1-5I_{\rm uni}/2)^{-2/3}-1\simeq-0.3.\label{beta}
\end{equation}
Equation (\ref{beta}) gives the $T=0$ energy correction in the universal limit when 
all two-particle correlations are evaluated using the ladder approximation. 
It should be noted that the medium effects giving 
energy shifts and Pauli blocking through $\Pi$ are significant; if the vacuum values  
$\rm{Im}\Pi_{\rm vac}=g^2mq/4\pi$ and $\tilde{\Pi}=0$ are  used in Eq.\ (\ref{Thermo3}), 
we obtain  $I_{\rm uni\ vac}=5/8$ which gives $\beta\simeq-0.519$.

\begin{figure}
\centering
\epsfig{file=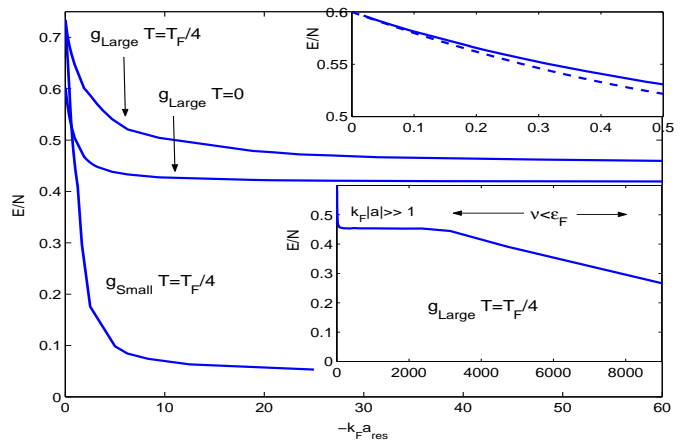,height=0.25\textheight,width=0.5\textwidth,angle=0}
\caption{The energy  per particle $E/N$ in units of $\epsilon_{\rm F}$ as a function of $-k_{\rm F}a_{\rm res}$
for a narrow and a broad resonance.}
\label{ENkfa}
\end{figure}

To illustrate the conclusions described above, we present results based on a numerical evaluation of 
Eq.\ (\ref{Thermo3}). 
In Fig.\ \ref{ENkfa},
 we plot the energy per particle $E/N$ as a function of $-k_{\rm F}a_{\rm res}$ for 
two different coupling strengths: $E_{g_{\rm large}}\gg \epsilon_{\rm F}$  and 
$E_{g_{\rm small}}\ll \epsilon_{\rm F}$. For $g_{\rm large}$, we also plot the energy 
for two different temperatures $T=0$ and $T=T_F/4$. The number of particles is fixed while 
$a_{\rm res}=-g^2m/(8\pi\nu)$ is 
varied through varying $\nu$. 
We see that for $g_{\rm large}$, the energy 
per particle approaches a constant for $-k_{\rm F}a_{\rm res}\gg 1$. In agreement with the 
Eq.\ (\ref{beta}), we find $E/N\simeq 0.42\epsilon_{\rm F}$ for $-k_{\rm F}a_{\rm res}\gg 1$
and $T=0$.
For the narrow resonance  $g_{\rm small}$ on the other hand, 
the energy per particle steadily decreases with increasing $-k_{\rm F}a_{\rm res}$ and the 
gas does not reach the universal limit for $a_{\rm res}<0$. This is 
simply because when $-k_{\rm F}a_{\rm res}\gtrsim 1$, we already have $\nu<\epsilon_{\rm F}$ for 
such a small value of $g$, and the depletion of the Fermi sea into the molecular level is 
significant. Of course, the depletion of the Fermi sea for $\nu\lesssim \epsilon_{\rm F}$ 
also happens for $g_{\rm large}$. However, since this corresponds to 
$-k_{\rm F}a_{\rm res}\gtrsim g_{\rm large}^2m^2/{4\pi k_{\rm F}}$, there is a large 
region of parameter space $1\ll -k_{\rm F}a_{\rm res}\ll g_{\rm large}^2m^2/{4\pi k_{\rm F}}$
where the thermodynamics of the gas is universal with no molecules present. This is illustrated 
by the inset, which plots $E/N$ for $g_{\rm large}$ over a larger range of $-k_{\rm F}a_{\rm res}$.
In a second inset, we plot $E/N$ for $-k_{\rm F}a_{\rm res}\ll 1$. The solid line 
is the numerical results whereas the dashed line is the Galitskii result 
$E/N=\epsilon_{\rm F}[3/5+2k_{\rm F}a_{\rm res}/(3\pi)+4(11-2\log 2)/(35\pi^2)(k_{\rm F}a)^2+\ldots]$. 
We see that our theory reproduces the Galitskii result for $-k_{\rm F}a_{\rm res}\ll 1$ as it should. 

\begin{figure}
\centering
\epsfig{file=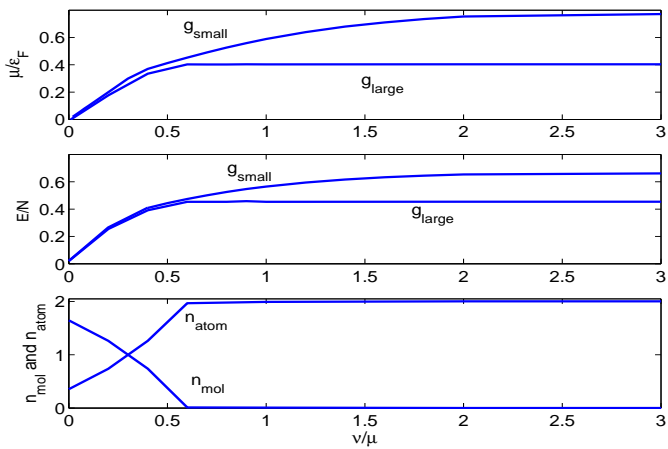,height=0.25\textheight,width=0.5\textwidth,angle=0}
\caption{The chemical potential $\mu$, the energy  per particle $E/N$ in units of $\epsilon_{\rm F}$ and 
the density of atoms and molecules in units of $n_\sigma=n/2=(2m\epsilon_{\rm F})^{3/2}/(6\pi^2)$ 
as a function of $\nu$ for a narrow and a broad resonance.}
\label{ENnu}
\end{figure}

The  depletion effect for $\nu\lesssim \epsilon_{\rm F}$
is illustrated more clearly in Fig.\ \ref{ENnu}, where we plot the 
chemical potential $\mu$ and $E/N$ for the same coupling strengths 
but now as function of $\nu$ for $T=T_F/4$.
We also plot the density  of (damped) molecules $n_{\rm mol}={\rm Tr}D$
and atoms $n_{\rm atom}=n-n_{\rm mol}$ for $g_{\rm large}$. We see that there is only a 
significant number of molecules 
when $\nu\lesssim \epsilon_F$ as expected. They will eventually become the 
undamped molecule gas for $\nu<0$. Of course, for both coupling strengths the 
gas converges toward a pure molecular state for $\nu<0$. 

In total, the numerical results confirm our conclusion, that the thermodynamics of the gas 
only exhibit universal behavior 
when $-a_{\rm res}k_{\rm F}\gg 1 $ for a sufficiently broad resonance with  
$E_{g_{\rm large}}\gg \epsilon_{\rm F}$. 

Let us now compare our result for the $T=0$ universal limit with 
a number of recent calculations. 
In ref.~\cite{Heiselberg}, the value $\beta\simeq -0.67$ was estimated from an expansion of 
the on-shell scattering amplitude in powers of $k_{\rm F}a$ and 
an approximate momentum average. 
In ref.~\cite{Baker}, the values $\beta\simeq-0.43$ and 
$\beta\simeq-0.67$ were suggested from two different Pad\'{e} approximants of the ladder  series 
for the energy. The $T=0$ result $\beta\simeq-0.3$
presented in this paper however, is the exact within the ladder approximation 
excluding superfluid correlations. A fixed node Greens function Monte Carlo calculation with $\sim 40$ particles 
yields  $\beta\simeq -0.46$ and $\beta\simeq-0.56$ when superfluid correlations are excluded and included 
respectively~\cite{Carlson}.  It could be interesting to 
examine in more detail the reasons for the difference between the present diagrammatic result and 
the Monte Carlo results in ref.~\cite{Carlson}. One possible reason could be  $>2$ body correlations
which cannot be expected to be small in the unitarity limit.

We conclude this paper by relating our results to some of the experiments on 
atomic gases. First, we examine whether a couple of the experimentally relevant resonances 
can be expected to exhibit universal thermodynamics. For 
one such Feshbach resonance at the magnetic field $B_0\simeq 200$ G
for $^{40}$K in the two open channel 
hyperfine states  $|9/2,-9/2\rangle$ and $|9/2,-7/2\rangle$~\cite{Bohn}, an 
analysis yields  $k_{\rm F}|a_{\rm res}|\simeq 15n_{12}^{-1/3}\epsilon_{\rm F}/\nu$
where $n_{12}$ is  the density in units of $10^{12}$ cm$^{-3}$~\cite{bruunpethick}.
Another case of interest is the $^6$Li Feshbach resonance between the two lowest hyperfine 
states at $B_0\simeq 860$ G~\cite{Houbiers}.
An estimate of this resonance yields 
$k_{\rm F}|a_{\rm res}| = 1.1\times10^4 n_{12}^{-1/3}  \epsilon_{\rm F}/\nu_{\rm bare}$~\cite{bruunpethick}.
This shows that these two particular resonances are sufficiently broad (large $g^2$) 
such that one would expect universal behavior on the $a_{\rm res}<0$ side of the resonance. 
There has recently been a number of measurements of the energy of a  $^6$Li gas close to the Feshbach 
resonance considered above~\cite{Duke,ENS,Innsbruck}. In agreement with the conclusions of 
this paper, they  suggest that the properties of the gas indeed become universal in the limit 
$|a|\rightarrow\infty$. The Duke group obtain $\beta=-0.26\pm 0.07$ 
on the $a_{\rm res}<0$ side of the $^6$Li 
 resonance~\cite{Duke}. This value agrees well with Eq.\ (\ref{beta}). 
However, the Innsbruck group measures $\beta=-0.68^{+0.13}_{-0.1}$ for the same resonance~\cite{Innsbruck}. 
Finally, the group at ENS obtains $(E_{\rm tot}-E_{\rm kin})/E_{\rm kin}\simeq-0.3$ 
for $a_{\rm res}\rightarrow-\infty$~\cite{ENS}. But since   
both the total energy $E_{\rm tot}$ and  the kinetic energy $E_{\rm kin}$ of the gas 
are measured for the interacting system this ratio is not identical to the parameter $\beta$ as 
defined in the present paper. 
In general, further investigation is needed in order to 
remove some of the apparent inconsistencies of the results for $|a|\rightarrow \infty$ reported so far.

In conclusion, we have shown that a two-component Fermi gas interacting via a Feshbach 
state can be expected to exhibit universal behavior  on the $a_{\rm res}<0$ side 
only when the resonance is sufficiently broad with $E_g\gg \epsilon_{\rm F}$.
We furthermore calculated the universal parameter $\beta$ taking into account two-body processes in 
the ladder approximation.
Our  results were compared to the experimental results when appropriate. 
\\

We acknowledge very useful discussions with C.\ J.\ Pethick, H.\ Heiselberg, and 
S.\ J.\ J.\ M.\ F.\ Kokkelmans.

\end{document}